\documentclass[%
aps,
pra,
longbibliography,
superscriptaddress,
amsmath,amssymb,
twocolumn,
]{revtex4-1}
\usepackage{mathrsfs}
\usepackage{graphicx}
\usepackage{amsfonts}
\usepackage{graphicx}
\usepackage{dcolumn}
\usepackage{bm}
\usepackage[colorlinks,urlcolor=blue,linkcolor=blue,citecolor=blue]{hyperref}
\usepackage{xcolor}

\begin{document}
	
\title{Controlling the superradiant phase transition in the quantum Rabi model}
\author{Xuan Xie}
\affiliation{Key Laboratory of Low-Dimensional Quantum Structures and Quantum Control of
	Ministry of Education, Key Laboratory for Matter Microstructure and Function of Hunan Province, Department of Physics and Synergetic Innovation Center for Quantum Effects and Applications, Hunan Normal University, Changsha 410081, China}

\author{Cheng Liu}
\affiliation{Key Laboratory of Low-Dimensional Quantum Structures and Quantum Control of
	Ministry of Education, Key Laboratory for Matter Microstructure and Function of Hunan Province, Department of Physics and Synergetic Innovation Center for Quantum Effects and Applications, Hunan Normal University, Changsha 410081, China}

\author{Lin-Lin Jiang}
\affiliation{Key Laboratory of Low-Dimensional Quantum Structures and Quantum Control of
	Ministry of Education, Key Laboratory for Matter Microstructure and Function of Hunan Province, Department of Physics and Synergetic Innovation Center for Quantum Effects and Applications, Hunan Normal University, Changsha 410081, China}

\author{Jin-Feng Huang}
\email{Corresponding author: jfhuang@hunnu.edu.cn}
\affiliation{Key Laboratory of Low-Dimensional Quantum Structures and Quantum Control of
	Ministry of Education, Key Laboratory for Matter Microstructure and Function of Hunan Province, Department of Physics and Synergetic Innovation Center for Quantum Effects and Applications, Hunan Normal University, Changsha 410081, China}
\affiliation{Institute of Interdisciplinary Studies, Hunan Normal University, Changsha 410081, China}

\begin{abstract}
	In the ultrastrong-coupling regime, the quantum Rabi model can exhibit quantum phase transition (QPT) when the ratio of the qubit transition frequency to the frequency of the cavity field approaches infinity. However, it is challenging to control the QPT in few-body systems because of the limited coupling strength and the $A^{2}$ term. Here, we propose a practical scheme to manipulate the QPT of quantum Rabi model in the strong-coupling regime. By applying a periodic frequency modulation to the two-level system in a quantum Rabi model in the strong-coupling regime, an anisotropic quantum Rabi model with ultrastrong and tunable coupling strengths for rotating and counter-rotating terms is obtained. The ground state and excitation energy of this model, in terms of the modulation parameters, are studied. We find that the QPT of the quantum Rabi model can be observed in the strong-coupling regime and externally controlled by the modulation.
\end{abstract}
\maketitle
\narrowtext
	
\section{Introduction}~\label{sectionI}
Quantum phase transition (QPT)~\cite{QPT1999} has long been an important and fundamental phenomenon in many-body systems, and has attracted a remarkable amount of study interest due to its special features. The QPT occurs when the non-thermal parameter, such as coupling strength, exceeds a critical value at zero temperature~\cite{QPT1999}. For example, the superradiant phase transition occurs in the Dicke model~\cite{Dicke} when the light-matter coupling strength exceeds a critical value in the thermodynamic limit, i.e., $N\rightarrow\infty$~\cite{1973dickle,1973spt,Sunchangpu,liyong,CEPRL,NLPRL,ABPRL,PRL120183603}. The control of the QPT in Dicke model was discussed~\cite{ABPRL,2023Modulation}. Recently, it was found that the QPT can also occur in few-body systems, such as the quantum Rabi model (QRM)~\cite{1936Rabi,1937Rabi} when the ratio of the qubit transition frequency $\omega_{0}$ to the frequency $\omega_{c}$ of the cavity field tends to infinity, i.e.,  $\eta=\omega_{0}/\omega_{c}\rightarrow\infty$~\cite{Hwang2015Rabi}. Hereafter, many efforts were devoted to studying the QPT in QRM both theoretically~\cite{MaoXinLiu2017,Rabi triangle,Rabi lattice,dispassive QRM,dar,sar,yingzujian,PRAaqrm,L. T. ShenPRA,full quantum limit} and experimentally~\cite{trapped ions,nuclear,ZhengShiBiao}, such as the universal scaling and critical exponents~\cite{MaoXinLiu2017}, QPT in the quantum Rabi triangle~\cite{Rabi triangle,Rabi lattice}, dissipative phase transition in the open QRM~\cite{dispassive QRM,dar,sar}, QPT in the anisotropic QRM~\cite{yingzujian,PRAaqrm} and quench dynamics in the anisotropic QRM~\cite{L. T. ShenPRA}. Even the QPT in QRM in the fully quantum limit has also been predicted~\cite{full quantum limit}.

Though the QPT in QRM was experimentally observed in a single trapped ion~\cite{trapped ions} and a nuclear magnetic resonance quantum simulator~\cite{nuclear}, it is still challenging to manipulate the ground-state QPT due to the limited coupling strength with the classical field (dispersion) limit $\eta\rightarrow\infty$ and the $A^{2}$ term in cavity quantum electrodynamics (QED)~\cite{A2 term,Hwang2015Rabi,no-go theorm}. The $A^{2}$ term can prevent the occurrence of the QPT in QRM~\cite{Hwang2015Rabi}, which can not be neglected in the ultrastrong-coupling regime in cavity-QED system.

In this work, we propose a practical scheme to manipulate the QPT in QRM. Concretely, a periodic modulation is applied to the qubit transition frequency in a standard QRM~\cite{2017manipulation,C. Liu} in the strong-coupling regime. Then we obtain an anisotropic QRM with tunable rotating and counter-rotating coupling strengths. Through the modulation, both the effective rotating and counter-rotating (CR) coupling strengths can be varied in a wide range and reach the ultrastrong-coupling regime. The ground state is allowed to be in the superradiant phase. To investigate the QPT of the anisotropic QRM, we calculate excitation energy and the ground-state phase in both the normal and superradiant phase in the entire coupling parameter space. We find that both the normal and superradiant phase can be reached in the strong-coupling regime via the modulation. We also find that Goldstone modes~\cite{Hwang2016JC,Goldstone} exist in the superradiant phase. The dependence of excitation energy and ground-state phase on the modulation parameters is analyzed. The feature of the QPT is also investigated by calculating the derivatives of the ground-state energy. Our results show that the QPT in the QRM can be observed and manipulated externally in the strong-coupling regime. These results may inspire other studies on QPT engineering in other few-body systems.

The rest of this paper is organized as follows. In Sec.~\ref{sectionII}, we introduce a qubit frequency modulation in a standard QRM and obtain an effective anisotropic QRM whose rotating and counter-rotating terms are tunable. In Sec.~\ref{sectionIII}, we analyze the dependence of the effective frequency ratio and the effective coupling strengths on modulation parameters. We obtain the excitation energies for five phases and the phase boundary in Sec.~\ref{section4}. In Sec.~\ref{section5}, we figure out the phase diagram and verify it by analyzing the order parameters. The feature of QPT is also discussed. We also discuss how to manipulate the ground-state phase by varying the modulation parameters in Sec.~\ref{section6}. In Sec.~\ref{section7}, we discuss the $A^{2}$ term under modulation. We present some discussions concerning the experimental implementation in Sec.~\ref{section8}. A brief conclusion is given in Sec.~\ref{section9}. Finally, three Appendixes are presented to show the validity evaluation of the rotating-wave approximation, the operators for superradiant phase, and the squeezing parameters.
\vspace{3pt}
\begin{figure}
	\begin{centering}
		\includegraphics[width=0.49 \textwidth]{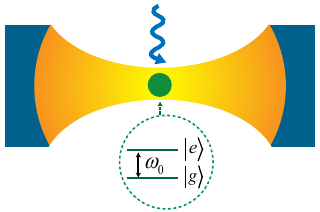}
		\par\end{centering}
	\centering{}\caption{(Color online) Schematic of a cavity field coupled to a two-level system (qubit) with a periodic modulation on the qubit transition frequency. }\label{Fig1}
\end{figure}
	
\section{Model and Hamiltonian}\label{sectionII}
The system we consider is composed of a two-level system (qubit) coupled to a cavity field, as shown in Fig.~\ref{Fig1}, which is described by the quantum Rabi model~\cite{1936Rabi,1937Rabi}. The corresponding Hamiltonian reads
\begin{equation}
	H_{\text{R}}=H_{\mathrm{\text{JC}}}+H_{\mathrm{\text{CR}}},
\end{equation}
where
\begin{equation}
	H_{\mathrm{\text{JC}}}=\omega_{c}a^{\dagger}a+\frac{\omega_{0}}{2}\sigma_{z}+g(\sigma_{+}a+a^{\dagger}\sigma_{-})
\end{equation}
is the Jaynes-Cummings (JC) Hamiltonian, and
\begin{equation}
	H_{\text{CR}}=g(\sigma_{-}a+a^{\dagger}\sigma_{+})
\end{equation}
describes the CR interactions. Here, $\sigma_{\pm}=(\sigma_{x}\pm i\sigma_{y})$/2 with $\sigma_{x,y,z}$ being the Pauli matrices, and $a\ (a^{\dagger})$ is the annihilation (creation) operator of the cavity field with frequency $\omega_{c}$. We use $\left|e\right\rangle$ and $\left|g\right\rangle$ to label the excited and ground states of the qubit, respectively. The transition frequency of the qubit is $\omega_{0}$ and
the coupling strength between the qubit and the cavity field is $g$. Note that in this work we use the terms of the cavity-QED system to expound the physical idea. Actually, the physical results also work for other physical platforms, in which the physical model consisting of a two-level system and a single-mode bosonic field can be implemented.
	
To engineer the system, we introduce a sinusoidal modulation to the transition frequency of the qubit. The
modulation Hamiltonian is given by~\cite{2017manipulation} 
\begin{equation}
	H_{\text{M}}(t)=\frac{1}{2}\xi\nu\cos(\nu t)\sigma_{z},
\end{equation}
where $\xi$ and $\nu$ are the dimensionless modulation amplitude and frequency, respectively. After including the two-photon $A^{2}$ term 
\begin{equation}
	H_{A^{2}}=g_{A^{2}}(a+a^{\dagger})^{2}\label{eq:0}
\end{equation}
with amplitude $g_{A^{2}}=\chi g^{2}/\omega_{0}$, where $\chi$ is a dimensionless coefficient, the total Hamiltonian can be expressed as
\begin{equation}
	H(t)=H_\text{R}+H_{A^{2}}+H_\text{M}(t).
\end{equation}
To obtain an effective Hamiltonian, we introduce
\begin{equation}
	H_{0}(t)=\omega_{c}^{\prime}a^{\dagger}a+\frac{1}{2}[\omega_{0}+\xi\nu\cos(\nu t)]\sigma_{z}
\end{equation}
with $\omega_{c}^{\prime}=\omega_{c}+2g_{A^{2}}$.
	
In the first rotating frame defined by
\begin{align}
	V_{1}(t) & =\exp\left\{-i\left[\left(\omega_{c}^{\prime}a^{\dagger}a+\frac{1}{2}\omega_{0}\sigma_{z}\right)t+\frac{1}{2}\xi\sin(\nu t)\sigma_{z}\right]\right\},
\end{align}
the Hamiltonian becomes
\begin{align}
	\tilde{H}(t)=&V_{1}^{\dagger}(t)H(t)V_{1}(t)-iV_{1}^{\dagger}(t)\dot{V}_{1}(t)\nonumber \\
	=&\sum_{n=-\infty}^{\infty}gJ_{n}(\xi)\sigma_{+}ae^{i\delta_{n}t}+\sum_{m=-\infty}^{\infty}gJ_{m}(\xi)\sigma_{+}a^{\dagger}e^{i\Delta_{m}t}\nonumber \\
	&+g_{A^{2}}a^{2}e^{-2i\omega_{c}^{\prime}t}+\text{H.c.},\label{eq:99}
\end{align}
where we omit the constant term in $H_{A^{2}}$. Here, $\delta_{n}=\omega_{0}-\omega_{c}^{\prime}+n\nu$ is the oscillating frequency for the $n$th sideband in the rotating terms, and $\Delta_{m}=\omega_{0}+\omega_{c}^{\prime}+m\nu$ is the oscillating frequency for the $m$th sideband in the CR terms. 
	
In the derivation of $\tilde{H}(t)$, we use the Jacobi-Anger expansion
\begin{equation}
	e^{i\xi\sin(\nu t)}=\sum_{n=-\infty}^{\infty}J_{n}(\xi)e^{in\nu t},
\end{equation}
where $J_{n}(\xi)$ is the $n$th Bessel function of the first kind. In Eq.~(\ref{eq:99}), a rotating sideband $n_{0}$ and a CR sideband $m_{0}$ satisfying 
$\left|gJ_{n_{0}}(\xi)/\delta_{n_{0}}\right|=\max\{\left|gJ_{n}(\xi)/\delta_{n}\right|,n\in Z\}$ and
$\left|gJ_{m_{0}}(\xi)/\Delta_{m_{0}}\right|=\max\{\left|gJ_{m}(\xi)/\Delta_{m}\right|,m\in Z\}$  
can be selected. Here, $\text{\textquotedblleft Z\textquotedblright}$ denotes all integers. The oscillating frequencies of all rotating and CR sidebands can be written as $\delta_{n_{0}+s}=\delta_{n_{0}}+s\nu$ and $\Delta_{m_{0}+s}=\Delta_{m_{0}}+s\nu$ with $s$ an integer, respectively. Thus the other sidebands are separated from the $n_{0}$ ($m_{0}$) sideband by $s\nu$. Under the conditions
\begin{equation}
	\nu\gg g>g\left|J_{m}(\xi)\right|,\ g\left|J_{n}(\xi)\right|\ \text{and}\ \nu\gg\left|\Delta_{m_{0}}\right|,\ \left|\delta_{n_{0}}\right|,\label{eq:100}
\end{equation} 
all the sidebands other than the $n_{0}$($m_{0}$)th sideband can be discarded using the rotating-wave
approximation (RWA). With $ g_{A^{2}}\ll g\ll\omega_{0}$ and $g_{A^{2}}\ll2\omega_{c}^{\prime}$, the $a^{2}$ and $a^{\dagger2}$ terms with oscillating frequencies $\pm2\omega_{c}^{\prime}$ from the two-photon $A^{2}$ Hamiltonian can also be neglected using the RWA. To be precise, the validity of this approximation is discussed in detail in Sec.~\ref{section7}. 
	
The Hamiltonian $\tilde{H}(t)$ then can be approximated as
\begin{equation}
	\tilde{H}_{1}(t)= g_{r}a\sigma_{+}e^{i\delta_{n_{0}} t}+g_{cr}\sigma_{+}a^{\dagger}e^{i\Delta_{m_{0}}t}+\text{H.c.},\label{eq:4}
\end{equation}
where we introduce the effective coupling strengths
\begin{equation}
	g_{r}(\xi)=gJ_{n_{0}}(\xi),\hspace{0.5cm} g_{cr}(\xi)=gJ_{m_{0}}(\xi).
\end{equation}
To evaluate the validity of the RWA made in Hamiltonian (\ref{eq:99}), we examine the dynamics of the fidelity between the evolving states governed by Hamiltonians (\ref{eq:99}) and (\ref{eq:4}). The results are shown in Appendix~\ref{appC}. The Hamiltonian describes an effective standard QRM with $g_{r}=g_{cr}$ or an effective anisotropic QRM with $g_{r}\neq g_{cr}$~\cite{QTxie} in interaction picture. Note that, we can tune the effective rotating and CR coupling strengths $g_{r}(\xi)$ and $g_{cr}(\xi)$ via the dimensionless modulation amplitude $\xi$. 
	
To make Hamiltonian (\ref{eq:4}) time independent, we introduce the effective frequencies 
\begin{equation}
	\tilde{\omega}_{0}(\nu) =\frac{\Delta_{m_{0}}+\delta_{n_{0}}}{2},\hspace{0.5cm}
	\tilde{\omega}_{c}(\nu)=\frac{\Delta_{m_{0}}-\delta_{n_{0}}}{2}
\end{equation}
for the qubit and the cavity field, respectively. Here, $\tilde{\omega}_{0}(\nu)$ and $\tilde{\omega}_{c}(\nu)$ depend on the modulation frequency $\nu$ via $\Delta_{m_{0}}$ and $\delta_{n_{0}}$. Hereafter, we denote $\tilde{\omega}_{0,c}(\nu)$ as $\tilde{\omega}_{0,c}$ for precision.
After transferring to the second rotating frame by the unitary transformation
	
\begin{equation}
	V_{2}(t)=e^{-i(\frac{\tilde{\omega}_{0}}{2}\sigma_{z}+\tilde{\omega}_{c}a^{\dagger}a)t},
\end{equation}
the Hamiltonian (\ref{eq:4}) becomes

\begin{align}
	{H}_{eff} & =V_{2}(t)\tilde{H}_{1}(t)V_{2}^{\dagger}(t)+i\dot{V}_{2}(t)V_{2}^{\dagger}(t)\nonumber \\
	&  =\frac{\tilde{\omega}_{0}}{2}\sigma_{z}+\tilde{\omega}_{c}a^{\dagger}a+g_{r}(a\sigma_{+}+\epsilon \sigma_{+}a^{\dagger}+\text{H.c.}).\label{eq:5}
\end{align}
Here, we introduce the anisotropic parameter $\epsilon(\xi)=g_{cr}(\xi)/g_{r}(\xi)$. The effective coupling strength $g_{r}(\xi)$, the effective frequencies $\tilde{\omega}_{0}(\nu)$ and $\tilde{\omega}_{c}(\nu)$, and the anisotropic parameter $\epsilon(\xi)$ can be  tuned by the modulation parameters $\nu$ and $\xi$. The effective Hamiltonian (\ref{eq:5}) describe an anisotropic QRM with $\epsilon(\xi)\neq1$ and a standard QRM with $\epsilon(\xi)=1$. Hereafter, we denote $\epsilon(\xi)$ as $\epsilon$ for precision.
\begin{figure}[t]
	\begin{centering}		\includegraphics[width=0.49 \textwidth]{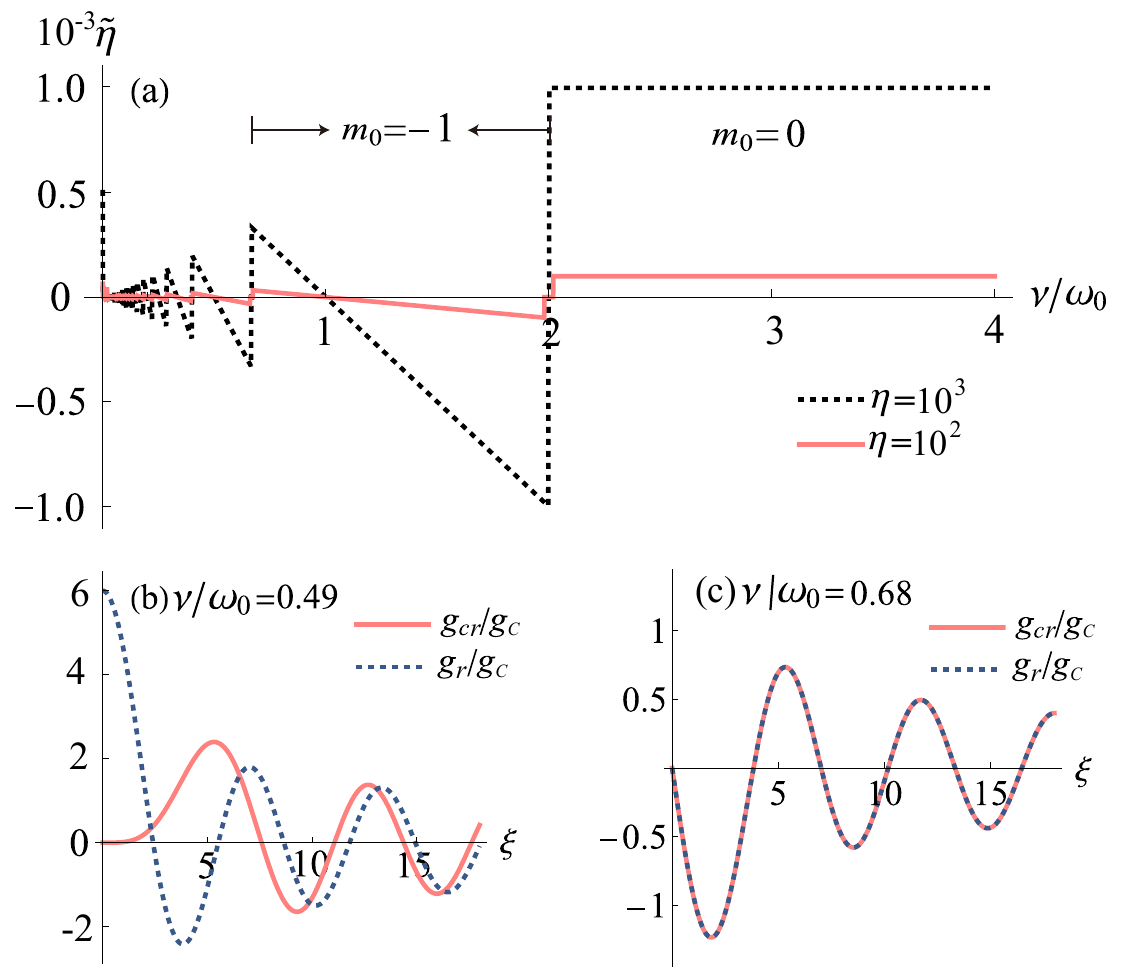}
		\par\end{centering}
	\centering{}\caption{(a) The effective ratio $\tilde{\eta}$ versus the modulation frequency $\nu/\omega_{0}$ for $\eta=10^{2}\text{ and }10^{3}.$ (b), (c) The couplings $g_{cr}/g_{C}$
		and $g_{r}/g_{C}$ versus the modulation amplitude $\xi$. Here, $\nu/\omega_{0}=0.49$ and $\eta=1$ in panel (b) and $\nu/\omega_{0}=0.68$ and $\eta=10^{2}$ in panel (c), the other parameter is $g/\omega_{0}=0.06.$}\label{Fig2}
\end{figure}
\section{Ultrastrong coupling}\label{sectionIII}
The effective QRM ($\epsilon=1$) in the limit $\tilde{\eta}=\tilde{\omega}_{0}/\tilde{\omega}_{c}\rightarrow\infty$ can exhibit quantum phase transition from the normal phase to the superradiant phase. This phase transition occurs at $g_{C}=\sqrt{\tilde{\omega}_{0}\tilde{\omega}_{c}}/2$~\cite{Ashhab2013} above which $Z_{2}$ parity symmetry is broken and the bosonic mode is macroscopically occupied~\cite{Hwang2015Rabi}. Here, the critical coupling strength $g_{C}$ depends on the modulation frequency $\nu$. Since the occurrence of the QPT in the quantum Rabi model requires the limit $\tilde{\eta}\rightarrow\infty$, to see clearly the dependence of $\tilde{\eta}$ on the modulation parameters, we plot $\tilde{\eta}$ versus $\nu$ for different ratios $\eta=\omega_{0}/\omega_{c}^{\prime}$ in Fig.~\ref{Fig2}(a). $\tilde{\eta}$ is independent of $\xi$. As shown in Fig.~\ref{Fig2}(a), the curve of the effective ratio $\tilde{\eta}$ is composed of a series of $\text{V}$-type valleys until $\nu$ increases up to $\nu/\omega_{0}=2(\omega_{0}+\omega_{c}^{\prime})$. The  $\text{V}$-shaped valleys share the same $m_{0}$ and the valley for $m_{0}$ starts at $\nu=2(\omega_{0}+\omega_{c}^{\prime})/(2\left|m_{0}\right|+1)$. When $\nu>2(\omega_{0}+\omega_{c}^{\prime})$, $m_{0}=0$ and $\tilde{\eta}=\eta$. Thus, in the regime $\nu<2(\omega_{0}+\omega_{c}^{\prime})$, we can manipulate $\tilde{\eta}$ via choosing appropriate $\nu$ while in the regime $\nu>2(\omega_{0}+\omega_{c}^{\prime})$, this $\tilde{\eta}$ cannot be varied via changing $\nu$. As shown in Fig.~\ref{Fig2}(a), to satisfy the condition (\ref{eq:100}) and the limit $\tilde{\eta}\rightarrow\infty$, the value of $\nu$ can be chosen in a broad range.
	
To see clearly the dependence of the effective coupling strengths $g_{r}(\xi)$ and $g_{cr}(\xi)$ on the modulation amplitude $\xi$, we plot $g_{cr}/g_{C}$ and $g_{r}/g_{C}$ versus the modulation amplitude $\xi$ for $\eta=1$ and $10^{2}$ at $\nu/\omega_{0}=0.49$ and $0.68$ in Figs.~\ref{Fig2}(b) and~\ref{Fig2}(c), respectively. The effective coupling strength $g_{r}/g_{C}\ ( g_{cr}/g_{C})$ depends on the modulation amplitude $\xi$ in the form of the Bessel function. When $\nu/\omega_{0}=0.49$ for $\eta=1$ $(m_{0}=-4,\ n_{0}=-1)$, $g_{r}/g_{C}$ oscillates between $-2.41656$ and 6 with reducing amplitude, and $g_{cr}/g_{C}$ oscillates between $-1.646$ and 2.398, as shown in Fig.~\ref{Fig2}(b).  When $\nu/\omega_{0}=0.68$ for $\eta=10^{2}$ $(m_{0}=n_{0}=-1)$, $g_{r}/g_{C}$ and $g_{cr}/g_{C}$ show the same oscillating feature between $-1.234$ and 0.734, as shown in Fig.~\ref{Fig2}(c). The results show that one can tune the effective coupling strengths $g_{r}$ and $g_{cr}$ in a broad range. 
\begin{figure}[t]
	\begin{centering}		\includegraphics[width=0.38 \textwidth]{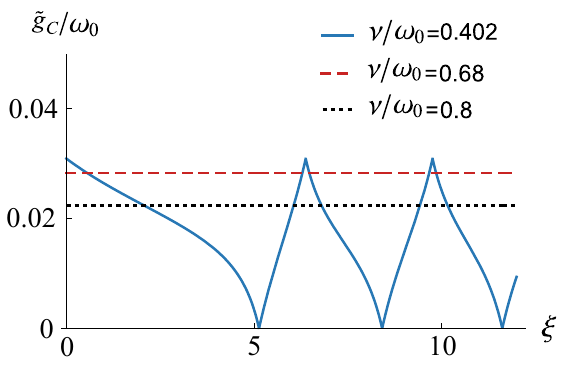}
		\par\end{centering}
	\centering{}\caption{The phase transition boundary $\tilde{g}_{C}/\omega_{0}$ versus $\xi$ for the anisotropic QRM at $\eta=10^{2}$ and $\nu/\omega_{0}=0.402$, $0.68$, and $0.8$.}\label{Fig3}
\end{figure}

\section{Excitation spectra}\label{section4}
To figure out the ground-state property of the system in the entire coupling-parameter space, we calculate the excitation energy in the limit $\tilde{\eta}\rightarrow\infty$ in this section.
\subsection{Normal phase}
To diagonalize the effective Hamiltonian (\ref{eq:5}), we introduce the canonical coordinate and momentum operators $x=(a+a^{\dagger})/\sqrt{2}$ and $p=i(a^{\dagger}-a)/\sqrt{2}$. In terms of $x$ and $p$, the Hamiltonian (\ref{eq:5}) reads
\begin{equation}
	H_\text{AR}=\tilde{H}_\text{0}+\tilde{H}_\text{I},\label{eq:2}
\end{equation}
where
\begin{subequations}
	\begin{align}
		\tilde{H}_{0}&=\frac{\tilde{\omega}_{c}}{2}(x^{2}+p^{2})+\frac{\tilde{\omega}_{0}}{2}\sigma_{z},\\
		\tilde{H}_\text{I}&=\frac{g_{r}}{\sqrt{2}}[(1+\epsilon)\sigma_{x}x-(1-\epsilon)\sigma_{y}p].
	\end{align}
\end{subequations} 
Then, in the limit $\tilde{\eta}\rightarrow\infty$, we make a unitary transformation $U_\text{N}^{\dagger}{H}_\text{AR}U_\text{N}$ on Hamiltonian (\ref{eq:2}) with
\begin{equation}
	U_\text{N}=\exp\{-ig_{r}[(1-\epsilon)\sigma_{x}p+(1+\epsilon)\sigma_{y}x]/{\sqrt{2}\tilde{\omega}_{0}}\}.\label{eq:19}
\end{equation}  
The transformed Hamiltonian becomes free of coupling between qubit
subspaces $\mathcal{H}_{g}$ and $\mathcal{H}_{e}$, which are spanned by $\left|g\right\rangle$ and $\left|e\right\rangle$, respectively. After a projection into $\mathcal{H}_{g}$, i.e., $\tilde{H}_\text{N}=\langle g|U_\text{N}^{\dagger}H_\text{AR}U_\text{N}|g\rangle$, we obtain an effective low-energy Hamiltonian
\begin{equation}
	\tilde{H}_\text{N}=\frac{{\tilde{\omega}_{c}}}{2}[(1-\zeta^{2})x^{2}+(1-\zeta^{\prime^{2}})p^{2}],\label{eq:24}
\end{equation} 
upto a constant term, where
\begin{equation}
	\zeta=\frac{\lambda(1+\epsilon)}{2},\hspace{0.5cm} \zeta^{\prime}=\frac{\lambda(1-\epsilon)}{2},
\end{equation} 
with the dimensionless rotating-coupling strength $\lambda\equiv{g_{r}/g_{C}}$. We note that there is no interaction term in Eq.~(\ref{eq:24}). From Eq.~(\ref{eq:24}), The excitation energy $\omega_\text{N}$ can be directly obtained as
\begin{equation}
	\omega_\text{N}=\tilde{\omega}_{c}\sqrt{(1-\zeta^{2})(1-\zeta^{\prime2})}.\label{eq:21}
\end{equation}

$\tilde{H}_\text{N}$ implies divergent values of $\langle x\rangle$ ($\langle p\rangle$) due to the disappearance of $x^{2}$ ($p^{2}$) term when $\epsilon>0$ ($\epsilon<0$) in the limit $\tilde{\eta}\rightarrow\infty$. Due to this fact, the normal phase is restricted in the region  

\begin{equation}
	g_{r}<\left|\tilde{g}_{C}\right|
\end{equation}
with 
\begin{equation}
	\tilde{g}_{C}=2g_{C}/(1+\left|\epsilon\right|).
\end{equation}
In the normal phase, the value of $\langle x\rangle\ (\langle p\rangle)$ is zero. $\tilde{g}_{C}$ determines the phase transition boundary for the anisotropic QRM. Note that, $g_{C}$ is the critical coupling strength for the standard QRM. When $g_{r}<\left|\tilde{g}_{C}\right|$, the ground state is in the normal phase with the cavity field in a vacuum and qubit in $\left|g\right\rangle$. While outside the normal phase, namely, $g_{r}>\left|\tilde{g}_{C}\right|$, the system is in the superradiant phase with both the qubit and the cavity field being excited.

In Fig.~\ref{Fig3}, we plot $\tilde{g}_{C}/\omega_{0}$ versus $\xi$ for $\nu/\omega_{0}=0.402$, $0.68$, and $0.8$. For $\nu/\omega_{0}=0.402$, $\tilde{g}_{C}/\omega_{0}$ oscillates between 0 and 0.031. The zero points of $\tilde{g}_{C}/\omega_{0}$ are located at $\xi=5.136$, $8.417$, and $11.62$, corresponding to the zeros of $J_{n_{0}}(\xi)$ ($n_{0}=-2$). For $(1+1/\eta)/(\left|n\right|+3/2)\leqslant\nu/\omega_{0}<(1-1/\eta)/(\left|n\right|+1/2)$ with $n\in\left\{ -N\right\}$ and $\nu/\omega_{0}\geq2(1+1/\eta)$, $\tilde{g}_{C}/\omega_{0}$ becomes independent of $\xi$ due to $m_{0}=n_{0}$. For example, when $\eta=10^{2}$ and $n=0$, the value of $\nu/\omega_{0}$ is between 0.673 and 1.98, such as $\nu/\omega_{0}=0.68$ and $0.8$ shown in Fig.~\ref{Fig3}. When $m_{0}=n_{0}$, the QPT boundary $\tilde{g}_{C}$ of the anisotropic QRM  returns back to the critical point ${g}_{C}$ of the QRM. Thus, the QPT boundary can be tuned via $\nu$ and $\xi$.
\vspace{-4.8pt}
\subsection{Superradiant phase}
Now, we study the excitation energy in the superradiant phase, namely, $g_{r}>\left|\tilde{g}_{C}\right|$. In the superradiant phase, there are four different phases. When $\epsilon>0$, the superradiant phase can be defined as superradiant $X$ (SX) phase since the $\langle x\rangle$ becomes divergent. When $\epsilon<0$, the system is in superradiant $P$ (SP) phase since the $\langle p\rangle$ is divergent. For $g_{cr}>2\left|g_{C}\right|$ with $g_{r}=0$, the ground state of system is in superradiant XP $a$ (SXPa) phase while for $g_{r}>2\left|g_{C}\right|$ with $g_{cr}=0$, the system is in superradiant XP $b$ (SXPb) phase.  When $g_{cr}=0$, only rotating terms exist and the Hamiltonian reduces to a Jaynes-Cummings model. In the following, we investigate the excitation energy in the limit $\tilde{\eta}\rightarrow\infty$ for SX, SP, SXPa, and SXPb phase, respectively. 
	
To effectively describe the superradiant phase with macroscopic excitation, we displace the cavity field $a$ by the displacing operator $D_{k}(\alpha)$ ($k$=SX, SP, SXPa, SXPb) for the $k$ phase. After the displacement, the Hamiltonian $H_\text{AR}$ becomes 
\begin{equation}
	\tilde{H}_{AR}=D_{k}^{\dagger}(\alpha)H_\text{AR}D_{k}(\alpha).
\end{equation}
Here, the detailed form of $D_{k}(\alpha)$ is given in the Appendix~\ref{appA}. To diagonalize the qubit part in Hamiltonian $\tilde{H}_{AR}$, we introduce a new set of Pauli operators $\tau_{x,y,z}^{k}$ ($k$=SX, SP, SXPa, SXPb) for each k phase, respectively. To be concise, the form of $\tau_{x,y,z}^{k}$ for the $k$ phase is also given in the Appendix~\ref{appA}. In terms of the new Pauli operators $\tau_{x,y,z}^{k}$ , the displaced Hamiltonian $\tilde{H}_{\text{AR}}$ becomes
\begin{equation}
	\begin{aligned}
		\tilde{H}_{\text{AR}}(\alpha_{\text{SX}})=&H_{c}+\frac{\Omega_{\text{SX}}}{2}\tau_{z}^{_{\text{SX}}}-\chi_{2}(\zeta^{\prime})\tau_{y}^{_{\text{SX}}}p+\chi_{1}(\zeta)\tau_{x}^{_{\text{SX}}}x \\
  &+\frac{\tilde{\omega}_{c}}{2}\alpha_{\text{SX}}^{2},\ \ \ \ \ \ \\
		\tilde{H}_{\text{AR}}(\alpha_{\text{SP}})=&H_{c}+\frac{\Omega_{\text{SP}}}{2}\tau_{z}^{_{\text{SP}}}-\chi_{1}(\zeta^{\prime})\tau_{x}^{_{\text{SP}}}p-\chi_{2}(\zeta)\tau_{y}^{_{\text{SP}}}x \\
		&+\frac{\tilde{\omega}_{c}}{2}\alpha_{\text{SP}}^{2},\ \ \ \ \ \ \ \ \\
		\tilde{H}_{\text{AR}}(\alpha_{\text{SXPa}})=&\frac{\Omega_{\text{SXPa}}}{2}\tau_{z}^{_{\text{SXPa}}}+\chi_{3}\left(\frac{4}{\mu}\tau_{x}^{_{\text{SXPa}}}x+\mu\tau_{y}^{_{\text{SXPa}}}p\right) \\
		&+H_{c}+\tilde{\omega}_{c}\alpha_{\text{SXPa}}^{2},\\
		\tilde{H}_{\text{AR}}(\alpha_{\text{SXPb}})=&\frac{\Omega_{\text{SXPb}}}{2}\tau_{z}^{_{\text{SXPb}}}+\chi_{3}\left(\frac{4}{\lambda}\tau_{x}^{_{\text{SXPb}}}x-\lambda\tau_{y}^{_{\text{SXPb}}}p\right) \\
		&+H_{c}+\tilde{\omega}_{c}\alpha_{\text{SXPb}}^{2}.
	\end{aligned}\label{eq:20}
\end{equation}
Here, the free cavity field Hamiltonian $H_{c}$ is 
\begin{equation}
	H_{c}=\frac{\tilde{\omega}_{c}}{2}(x^{2}+p^{2})-\frac{\tilde{\omega}_{c}}{2},
\end{equation}
and the rescaled qubit frequency $\Omega_{k}$ for the $k$ phase is
\begin{equation}
	\begin{array}{r@{}l}
		\Omega_{\text{SX}}& {}=\tilde{\omega}_{0}\zeta^{2},\ \\
		\Omega_{\text{SP}}& {}=\tilde{\omega}_{0}\zeta^{\prime2},\\
		\Omega_{\text{SXPa}}& {}=g_{cr}^{2}/\tilde{\omega}_{c},\ \\
		\Omega_{\text{SXPb}}& {}=g_{r}^{2}/\tilde{\omega}_{c}.\ \ 
	\end{array}
\end{equation}
The displacement parameter $\alpha$ is determined by the vanishing of the non-diagonal terms during deriving Eq. (\ref{eq:20}), which is obtained as
\begin{equation}
	\begin{aligned}
		\alpha_{\text{SX}}&=\pm\sqrt{\frac{\tilde{\eta}}{2}\left(\zeta^{2}-\frac{1}{\zeta^{2}}\right)}, \ \ \ \ \\
		\alpha_{\text{SP}}&=\pm\sqrt{\frac{\tilde{\eta}}{2}\left(\zeta^{\prime2}-\frac{1}{\zeta^{\prime2}}\right)},\ \ \ \\
		\alpha_{\text{SXPa}}&=\pm\sqrt{\tilde{\eta}\left(\frac{\mu^{2}}{16}-\frac{1}{\mu^{2}}\right)},\\
		\alpha_{\text{SXPb}}&=\pm\sqrt{\tilde{\eta}\left(\frac{\lambda^{2}}{16}-\frac{1}{\lambda^{2}}\right)},
	\end{aligned}
\end{equation}
for the $k$ ($k$=SX, SP, SXPa, SXPb) phase. To distinguish $\alpha$ in each $k$ phase, here we add a subscript $k$ in $\alpha$, i.e., we label $\alpha$ as $\alpha_{k}$. Here, the dimensionless CR coupling strength $\mu$ is defined as $\mu=g_{cr}/g_{C}$. The parameter $\chi_{i}$ ($i$=1,2,3) in Eq. (\ref{eq:20}) is defined as
\begin{equation}
	\chi_{1}(\zeta)=\frac{\sqrt{2}g_{C}}{\zeta},\ \ \chi_{2}(\zeta)=\sqrt{2}g_{C}\zeta,\ \text{and}\ 
	\chi_{3}=\frac{g_{C}}{\sqrt{2}}.
\end{equation}
	
We find that the form of the transformed Hamiltonian $\tilde{H}_{\text{AR}}(\alpha_{k})$ in Eq. (\ref{eq:20}) is similar to Hamiltonian (\ref{eq:2}) with rescaled qubit frequency $\Omega_{k}$. 
\begin{figure}[t]
	\begin{centering}
		\includegraphics[width=0.49 \textwidth]{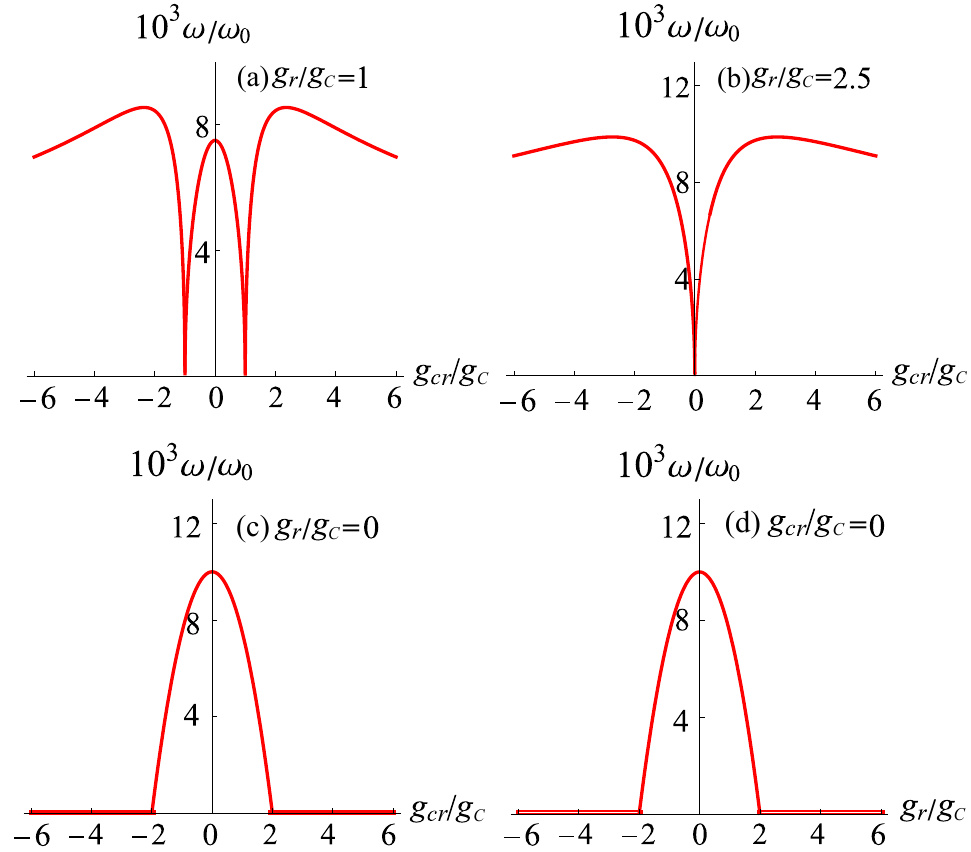}
		\par\end{centering}
	\centering{}\caption{(a)-(c) The excitation energy versus the coupling $g_{cr}/g_{C}$
		at $g_{r}/g_{C}=1,\ 2.5$, and $0$, respectively. (d) The excitation energy
		versus the coupling $g_{r}/g_{C}$ at $g_{cr}/g_{C}=0$. Other
		parameters are $\eta=10^{2}$ and $\nu/\omega_{0}=0.68$.}\label{Fig4}
\end{figure}
Thus, we can make a similar unitary transformation $U_{k}$ as $U_{\text{N}}$ in the normal phase on $\tilde{H}_\text{AR}(\alpha_{k})$ to eliminate the non-diagonal terms, namely,
\begin{equation}
	\tilde{H}_{k}=U_{k}^{\dagger}\tilde{H}_{\text{AR}}(\alpha_{k})U_{k},
\end{equation}($k$=SX, SP, SXPa, SXPb).
Here, the detailed form of $U_{k}$ is given in the Appendix~\ref{appA}. After performing a similar procedure used to derive $\tilde{H}_\text{N}$, we obtain the effective Hamiltonian $\tilde{H}_{k}$ in the $k$ phase as follows:
\begin{equation}
	\begin{aligned}
		\tilde{H}_{\text{SX}}&=\frac{\tilde{\omega}_{c}}{2}\left(1-\frac{1}{\zeta^{4}}\right)x^{2}+\frac{\tilde{\omega}_{c}}{2}\left(1-\frac{\zeta^{\prime2}}{\zeta^{2}}\right)p^{2},\ \ \ \ \ \ \ \ \ \ \ \ \ \ \ \ \ \ \ \ \ \ \\
		\tilde{H}_{\text{SP}}&=\frac{\tilde{\omega}_{c}}{2}\left(1-\frac{\zeta^{2}}{\zeta^{\prime2}}\right)x^{2}+\frac{\tilde{\omega}_{c}}{2}\left(1-\frac{1}{\zeta^{\prime4}}\right)p^{2},\ \ \ \ \ \ \ \ \ \ \ \ \ \ \ \ \ \ \ \ \ \ \\
		\tilde{H}_{\text{SXPa}}&=\frac{\tilde{\omega}_{c}}{4}\left(1-\frac{16}{\mu^{4}}\right)x^{2}-\frac{\tilde{\omega}_{c}}{2}\left(1+\frac{4}{\mu^{2}}\right)-\tilde{\omega}_{0}\left(\frac{\mu^{2}}{16}+\frac{1}{\mu^{2}}\right),\\
		\tilde{H}_{\text{SXPb}}&=\frac{\tilde{\omega}_{c}}{4}\left(1-\frac{16}{\lambda^{4}}\right)x^{2}-\frac{\tilde{\omega}_{c}}{2}\left(1-\frac{4}{\lambda^{2}}\right)-\tilde{\omega}_{0}\left(\frac{\lambda^{2}}{16}+\frac{1}{\lambda^{2}}\right).
	\end{aligned}\label{eq:35}
\end{equation}
Note that Hamiltonian $\tilde{H}_{k}$ in Eq. (\ref{eq:35}) describes a free bosonic mode. Interestingly, the $p^{2}$ term disappears in $\tilde{H}_{\text{SXPa}}$ and $\tilde{H}_{\text{SXPb}}$. Now, the excitation energy $\omega_{k}$ for each $k$ phase can be obtained as
\begin{equation}
	\begin{array}{r@{}l}
		\omega_{\text{SX}}& {}=\tilde{\omega}_{c}\sqrt{\left(1-\frac{1}{\zeta^{4}}\right)\left(1-\frac{\zeta^{\prime2}}{\zeta^{2}}\right)},\ \\
		\omega_{\text{SP}}& {}=\tilde{\omega}_{c}\sqrt{\left(1-\frac{1}{\zeta^{\prime4}}\right)\left(1-\frac{\zeta^{2}}{\zeta^{\prime2}}\right)},\\
		\omega_{\text{SXPa}}& {}=0,\ \ \ \ \ \ \ \ \ \ \ \ \ \ \ \ \ \ \ \ \ \ \ \ \ \ \ \ \ \ \ \ \ \ \ \ \\
		\omega_\text{SXPb}& {}=0.\ \ \ \ \ \ \ \ \ \ \ \ \ \ \ \ \ \ \ \ \ \ \ \ \ \ \ \ \ \ \ \ \ \ \ \  
	\end{array}\label{eq:1}
\end{equation}
\begin{table}[t]
	\renewcommand{\arraystretch}{1.3}
	\caption{$\langle x\rangle$, $\langle p\rangle$, $\langle x^{2}\rangle$, and $\langle p^{2}\rangle$ for all phases.}
	\label{table_one}
	\begin{ruledtabular}
		\begin{tabular*}{\linewidth}{cccccc}
			& N & SX & SP & SXPa & SXPb\\	
			\hline
			$\langle x\rangle$ & 0 & $s(\zeta)$ & 0 &$s(\frac{\mu}{2})$ & $s(\frac{\lambda}{2})$\\
			\hline
			$\langle p\rangle$ & 0 & 0 & $s(\zeta^{\prime})$ &0 & 0\\
			\hline	
			$\Delta x^{2}$ & $v(\zeta,\zeta^{\prime})$& $w(\epsilon,\zeta)$ & $4^{-1}w^{-1}(-\epsilon,\zeta^{\prime})$ & 0&0 \\
			\hline	
			$\Delta p^{2}$ & $v(\zeta^{\prime},\zeta)$& $4^{-1}w^{-1}(\epsilon,\zeta)$ & $w(-\epsilon,\zeta^{\prime})$ & $\infty$&$\infty$ \\
		\end{tabular*}
	\end{ruledtabular}
\end{table}
	
We find that the excitation energy closes both in the SXPa and SXPb phases. This result denotes the existence of the Goldstone modes~\cite{Hwang2016JC,Goldstone}. Furthermore, $\omega_\text{{SP}}$ has the same form as $\omega_\text{{SX}}$ if making the replacement $\text{\ensuremath{\zeta\leftrightarrow}\ensuremath{\zeta^{\prime}}}$ in $\omega_\text{{SX}}$.
\begin{figure}[b]
	\begin{centering}
		\includegraphics[width=0.49 \textwidth]{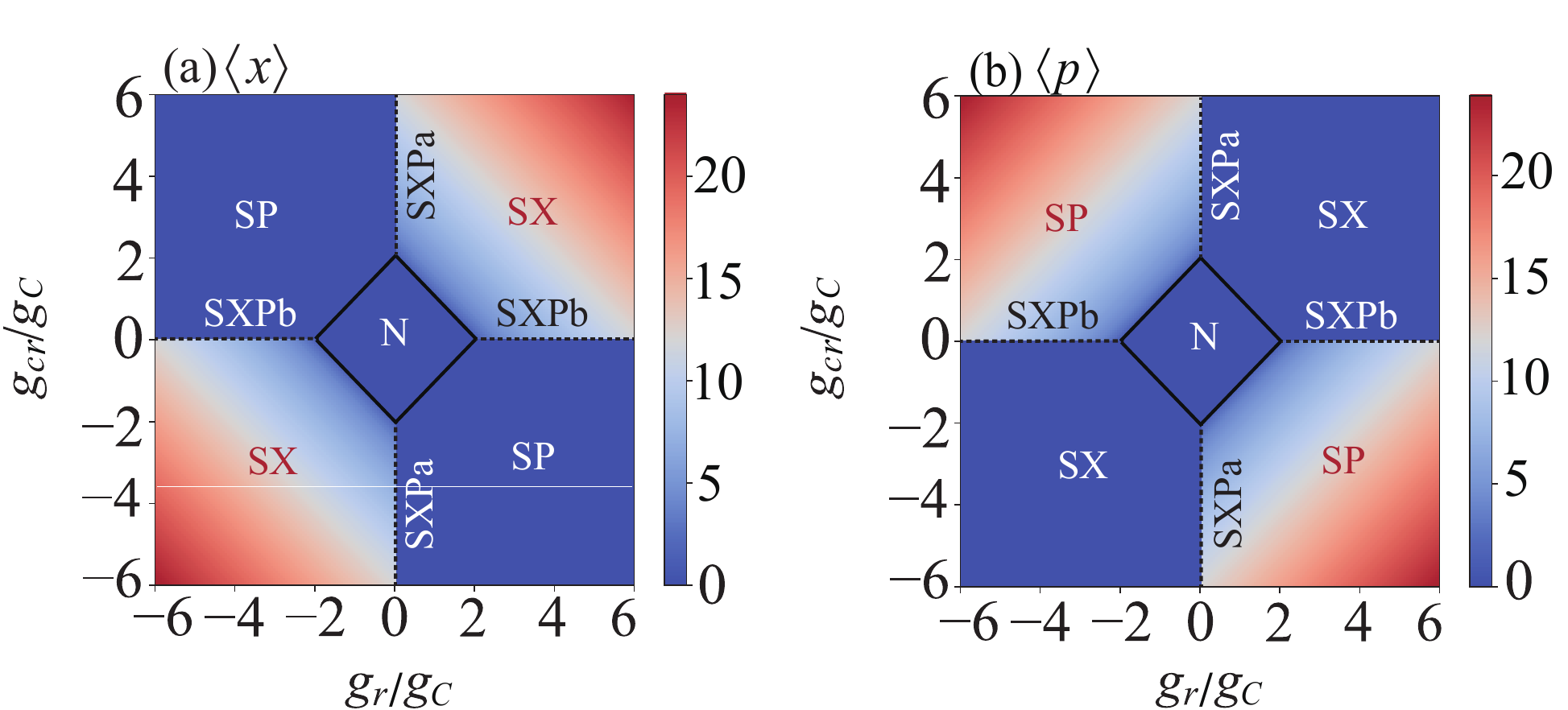}
		\par\end{centering}
	\centering{}\caption{$\langle x\rangle$ in panel (a) and $\langle p\rangle$ in panel (b) versus the coupling strengths $g_{r}/g_{C}$ and $g_{cr}/g_{C}$ for $\eta=10^{2}$.}\label{Fig6}
\end{figure}
	
To see clearly the feature of the excitation energy, we plot the excitation energy versus the coupling strength $g_{r}$ or $g_{cr}$ in Fig.~\ref{Fig4}.  For $g_{r}/g_{C}=1,$ as shown in Fig.~\ref{Fig4}(a), the critical points occur at $g_{cr}/g_{C}=1\ \text{and}-1$. Along the $x$-axis, the system experiences the phase transitions from the SP phase to the normal phase at $g_{cr}/g_{C}=-1$, and from the normal phase to the SX phase at $g_{cr}/g_{C}=1$. For $g_{r}/g_{C}=2.5$ as shown in Fig.~\ref{Fig4}(b), a single critical point emerges at $g_{cr}/g_{C}=0$. This point just denotes the SXPb phase. Along the $x$-axis, the system experiences phase transitions SP$\rightarrow$SXPb$\rightarrow$SX. The case of $g_{r}/g_{C}=0$ is shown in Fig.~\ref{Fig4}(c). As indicated by the curve in Fig.~\ref{Fig4}(c), for $\left|g_{cr}/g_{C}\right|>2$, the system reaches the SXPa phase where the U(1) symmetry is broken and the excitation becomes gapless. We also consider the case $g_{cr}/g_{C}=0$, as shown in Fig.~\ref{Fig4}(d). Similarly, for $g_{cr}/g_{C}=0$, the phase transitions between the normal phase and SXPb phase occur at $g_{r}/g_{C}=\pm2$. In the regime $\left|g_{r}/g_{C}\right|>2$, the system reaches the SXPb phase and the excitation energy is gapless. This result also denotes a Goldstone mode.
\begin{figure}[t]
	\begin{centering}
		\includegraphics[width=0.49 \textwidth]{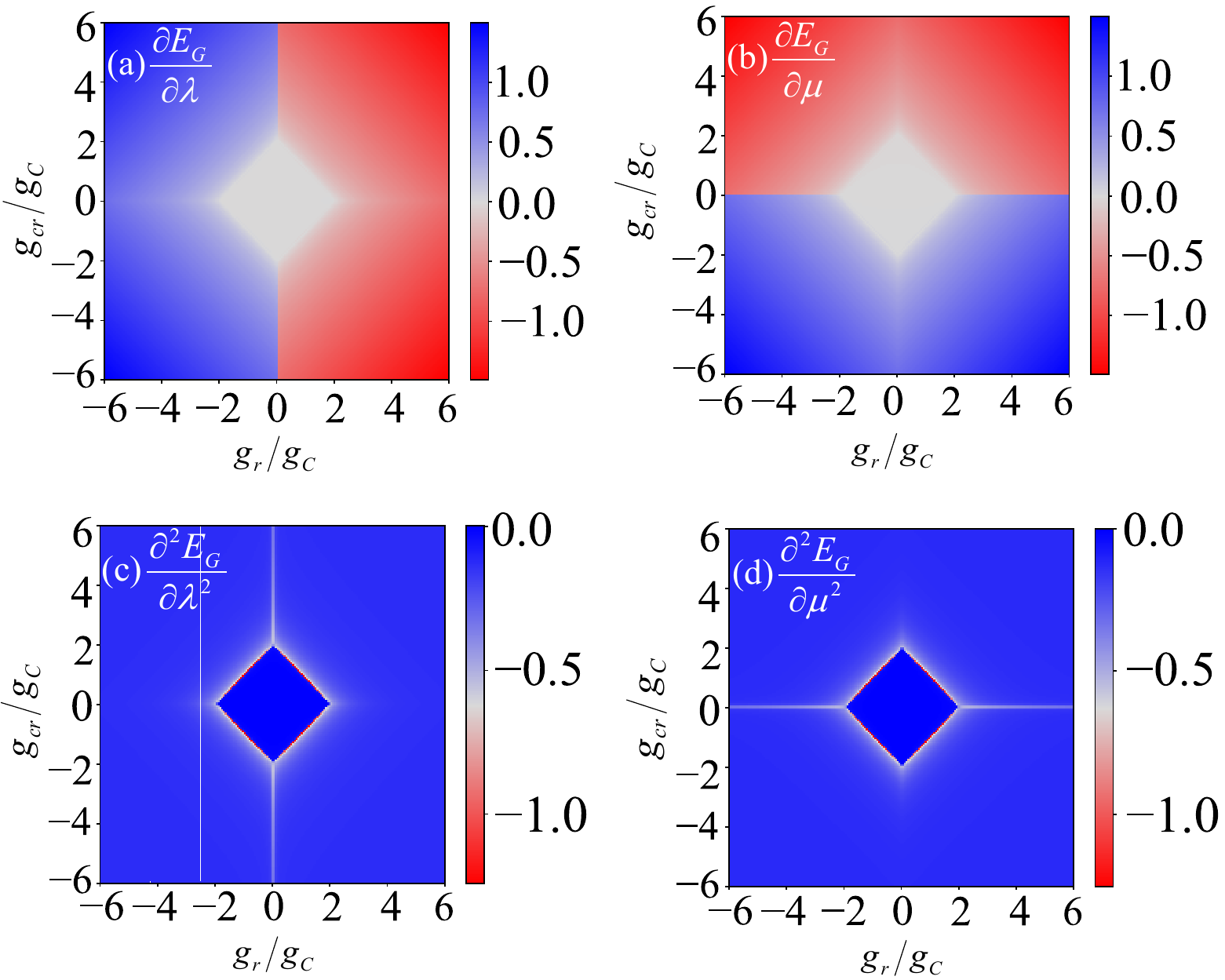}
		\par\end{centering}
	\centering{}\caption{(a), (b) The first derivatives $\partial E_{G}/\partial\lambda$ and $\partial E_{G}/\partial\mu$ versus the coupling strengths $g_{r}/g_{C}$ and $g_{cr}/g_{C}$.  (c), (d) The second derivatives $\partial^{2}E_{G}/\partial\lambda^{2}$ and $\partial^{2}E_{G}/\partial\mu^{2}$
		versus the coupling strengths $g_{r}/g_{C}$ and $g_{cr}/g_{C}$. Other parameters are $\omega_{0}=1,\eta=10^{2}$.}\label{Fig8}
\end{figure}	
\section{Phase diagram}\label{section5}
In this section, we investigate the phase diagram based on the above analysis and order parameters. Then we discuss the feature of QPTs by studying the discontinuity of the derivatives of the ground-state energy $E_{G}$.

To study the phase diagram, we introduce the average values $\langle x\rangle$ and $\langle p\rangle$ as the order parameters. The average values $\langle x\rangle$ and $\langle p\rangle$, and the corresponding variances $\Delta x^{2}$ and $\Delta p^{2}$ for $p$ phase are obtained from the ground state of the Hamiltonian $\tilde{H}_{p}$ ($p$=N, SX, SP, SXPa, SXPb). We show the results in Table \ref{table_one}. The expressions of $s$, $v$, and $w$ in Table \ref{table_one} are defined as 
\begin{figure}[t]
		\begin{centering}
			\includegraphics[width=0.38 \textwidth]{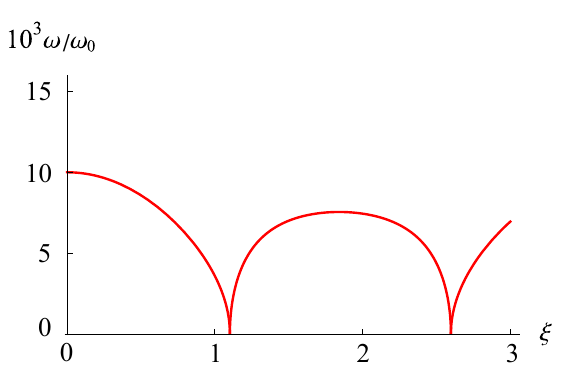}
			\par\end{centering}
		\centering{}\caption{(Color online) The excitation energy $\omega$ versus the dimensionless modulation amplitude $\xi$. Parameters
			are $\eta=10^{2},\nu/\omega_{0}=0.68$, and $g/\omega_{0}=0.06$.}\label{Fig5}
	\end{figure}
	\begin{subequations}
	\begin{align}
			s(x)&=\sqrt{\frac{\text{\ensuremath{\tilde{\eta}}}}{2}\left(x^{2}-\frac{1}{x^{2}}\right)},\\
			v(x_{1}, x_{2})&=\frac{1}{2}\sqrt{\frac{1-x_{2}^{2}}{1-x_{1}^{2}}},\\
			w(x_{1},x_{2})&=\frac{\sqrt{x_{1}}}{1+x_{1}}\frac{x_{2}^{2}}{\sqrt{x_{2}^{4}-1}}.
		\end{align}
	\end{subequations}
\begin{figure}[b]
	\begin{centering}
		\includegraphics[width=0.49 \textwidth]{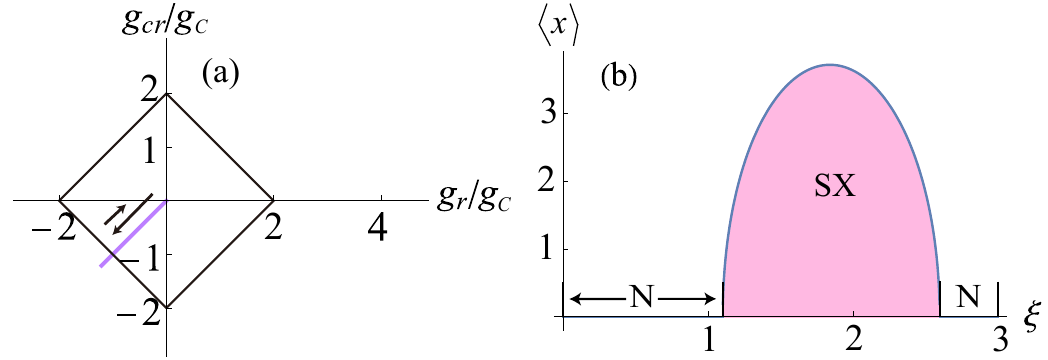}
		\par\end{centering}
	\centering{}\caption{(Color online) (a) The parameterized curve of the couplings $g_{cr}/g_{C}$ and $g_{r}/g_{C}$ with the modulation amplitude $\xi$ increasing from 0 to 3 as indicated by the arrows. (b) $\langle x\rangle$ versus the modulation amplitude $\xi$. The parameters are the same as Fig.~\ref{Fig5}.}\label{Fig7}
\end{figure}

It is found that, in the limit $\tilde{\eta}\rightarrow\infty$, we have $\langle x\rangle\rightarrow\infty$ and $\langle p\rangle=0$ in the SX phase while $\langle x\rangle=0$ and $\langle p\rangle\rightarrow\infty$ in the the SP phase. That is why they are defined as the SX phase and SP phase, respectively. We also note that the variances $\Delta x^{2}$ and $\Delta p^{2}$ are finite for N, SX, and SP phases, but $\Delta x^{2}=0$ and $\Delta p^{2}=\infty$ for both the SXPa and SXPb phases.
	
To see clearly the dependence of $\langle x\rangle$ and $\langle p\rangle$ on the coupling strengths $g_{r}$ and $g_{cr}$, we plot $\langle x\rangle$ and $\langle p\rangle$ versus the coupling strengths $g_{r}$ and $g_{cr}$ in Fig.~\ref{Fig6}. In the region $g_{r}<\left|\tilde{g}_{C}\right|$, both $\langle x\rangle$ and $\langle p\rangle$ are 0, which corresponds to the normal phase, as labeled by $N$. Outside the normal phase region, i.e., $g_{r}>\left|\tilde{g}_{C}\right|$, when $\epsilon>0$, the ground state is in the SX phase with $\langle x\rangle>0$; when $\epsilon<0$, the ground state is in the SP phase with $\langle p\rangle>0$. One can distinguish the phases by $\langle x\rangle$ and $\langle p\rangle$.

To characterize the type of the QPTs, we calculate the derivatives of ground-state energy $E_{G}$ over the coupling strengths $g_{r}$ and $g_{cr}$. To start with, we first calculate the ground-state energy $E_{G}$ in all phases. We perform the squeezing transformation 
\begin{equation}
	S(r_{p})=\exp\left[\frac{r_{p}}{2}(a^{2}-a^{\dagger2})\right]
\end{equation}
on $\tilde{H}_{p}$ ($p$=N, SX, SP, SXPa, and SXPb). The detailed form of $r_{p}$ is given in the Appendix~\ref{appB}. Then the resulting Hamiltonian reads
\begin{align}
	\tilde{H}(r_{p})=&S^{\dagger}(r_{p})\tilde{H}_{p}S(r_{p})\nonumber \\
	=&\omega_{p}a^{\dagger}a+E_{G}^{p},
\end{align}
where $\omega_{p}$ is the excitation energy given by Eqs. (\ref{eq:21}) and (\ref{eq:1}), and $E_{G}^{p}$ is the ground-state energy in $p$ phase, which is obtained as
\begin{equation}
	\begin{aligned}
		E_{G}^{\text{N}}=&\frac{\omega_{\text{N}}}{2}+\frac{\tilde{\omega}_{c}}{2}\zeta\zeta^{\prime}-\frac{\tilde{\omega}_{0}}{2}-\frac{\tilde{\omega}_{c}}{2},\\
		E_{G}^{\text{SX}}=&\frac{\omega_{\text{SX}}}{2}+\frac{\tilde{\omega}_{c}}{2}\frac{\zeta^{\prime}}{\zeta^{3}}-\frac{\tilde{\omega}_{0}}{4}(\zeta^{2}+\zeta^{-2})-\frac{\tilde{\omega}_{c}}{2},\\
		E_{G}^{\text{SP}}=&\frac{\omega_{\text{SP}}}{2}+\frac{\tilde{\omega}_{c}}{2}\frac{\zeta}{\zeta^{\prime3}}-\frac{\tilde{\omega}_{0}}{4}(\zeta^{\prime2}+\zeta^{\prime-2})-\frac{\tilde{\omega}_{c}}{2},\\
		E_{G}^{\text{SXPa}}=&-\frac{\tilde{\omega}_{c}}{2}\left(1+\frac{4}{\mu^{2}}\right)-\tilde{\omega}_{0}\left(\frac{\mu^{2}}{16}+\frac{1}{\mu^{2}}\right),\\
		E_{G}^{\text{SXPb}}=&-\frac{\tilde{\omega}_{c}}{2}\left(1-\frac{4}{\lambda^{2}}\right)-\tilde{\omega}_{0}\left(\frac{\lambda^{2}}{16}+\frac{1}{\lambda^{2}}\right).
	\end{aligned}	
\end{equation}
	
We plot the dependence of the first and second derivatives of $E_{G}$ on $\lambda$ and $\mu$ versus $g_{r}/g_{C}$ and $g_{cr}/g_{C}$ in Fig.~\ref{Fig8}. Note that, $\lambda$ and $\mu$ are dimensionless rotating and CR coupling strengths, respectively. As shown in Fig.~\ref{Fig8}(a) and \ref{Fig8}(b), it is found that $\partial E_{G}/\partial\lambda$ ($\partial E_{G}/\partial\mu$) is discontinuous at $g_{r}=0\ (g_{cr}=0)\ \text{for}\ \left|g_{cr}/g_{C}\right|>2$ ($\left|g_{r}/g_{C}\right|>2$). This result denotes the QPT from the SX phase to the SP phase is of first order. As shown in Figs.~\ref{Fig8}(c) and \ref{Fig8}(d), we observe that $\partial^{2}E_{G}/\partial\lambda^{2}$ ($\partial^{2}E_{G}/\partial\mu^{2}$) is discontinuous at the boundaries $g_{r}=\left|\tilde{g}_{C}\right|$ and $g_{r}=0$ ($g_{cr}=0$) for $\left|g_{cr}/g_{C}\right|>2$ ($\left|g_{r}/g_{C}\right|>2$). Thus, the transition from the normal phase to superradiant phases is of second order. We also use the dashed and solid lines to indicate the first- and second-order QPT in Fig.~\ref{Fig6}, respectively.
		
\section{Manipulation of Quantum phases}\label{section6}
In this section, we discuss how to manipulate the ground state by varying the modulation parameters. We plot the excitation energy $\omega$ as a function of $\xi$ at $\nu/\omega_{0}=0.68$ in Fig.~\ref{Fig5}. The subscript $p$ that labels the specific phase in $\omega_{\text{p}}$ is omitted. The solid curve indicates that the system can enter different phases by adjusting the dimensionless modulation amplitude $\xi$. The QPT occurs at $\xi=1.102$ and $2.598$, where $g_{r}/g_{C}=-1$, $g_{cr}/g_{C}=-1$. The ground state is in the normal phase when $0<\xi<1.102$ and $2.598<\xi<3$, and the SX phase appears when  $1.102<\xi<2.598$. The system experiences the QPTs from the normal phase to the SX phase at $\xi=1.102$ and from the SX phase to the normal phase at $\xi=2.598$.
\begin{figure}[t]
	\begin{centering}
		\includegraphics[width=0.49 \textwidth]{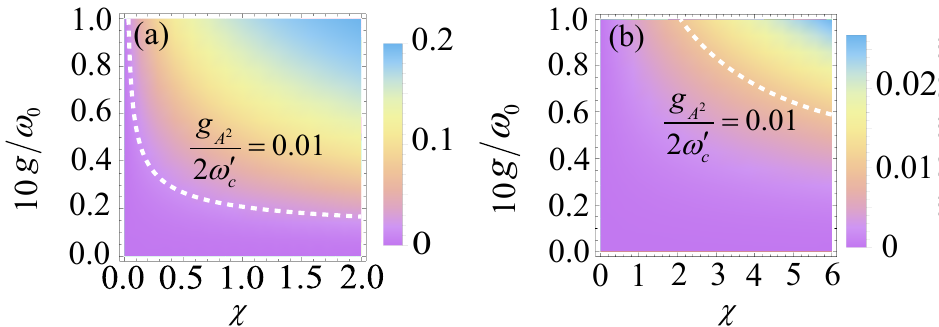}
		\par\end{centering}
	\centering{}\caption{(Color online) The ratio $g_{A^{2}}/2\omega_{c}^{\prime}$ versus the coupling strength $g/\omega_{0}$ and the dimensionless factor $\chi$. Parameters are (a) $\omega_{0}/\omega_{c}=10^{2}$ and (b) $\omega_{0}/\omega_{c}=1$.}\label{Fig10}
\end{figure}
	
To see the system how crosses the phase boundary under the chosen parameter in Fig.~\ref{Fig5}, we plot the parameterized curves of [$g_{r}(\xi)/g_{C}$, $g_{cr}(\xi)/g_{C}$] versus the modulation amplitude $\xi$ in Fig.~\ref{Fig7}(a). The parameterized curve evolves from the normal phase to the SX phase and then back to the normal phase as $\xi$ increases from 0 to 3. The system can be described by a standard QRM. In Fig.~\ref{Fig7}(b), we plot $\langle x\rangle$ versus the modulation amplitude $\xi$ for the same parameters as Fig.~\ref{Fig7}(a). Note that, $\langle p\rangle=0$ under the parameters in Fig.~\ref{Fig7}(a). As shown in Fig.~\ref{Fig7}(b), although the coupling strength is only in the strong-coupling regime with $g/\omega_{0}=0.06$, yet by the modulation, the QRM can reach the superradiant phase when $1.102<\xi<2.598$. The normal phase appears when $0<\xi<1.102$ and $2.598<\xi<3$. The results shown in Fig.~\ref{Fig7} are consistent with those in Fig.~\ref{Fig5}. Thus, we can manipulate quantum phases on the QRM via the appropriate modulation with strong coupling.

We mention that the anisotropic QRM can be realized with our scheme, as shown in Fig.~\ref{Fig2}(b). However, to reach the classical field limit $\tilde{\eta}\rightarrow\infty$, only an effective standard QRM can be realized. This is because in the limit $\tilde{\eta}\rightarrow\infty$ and the condition (\ref{eq:100}), we obtain $m_{0}=n_{0}$, thus $g_{r}=g_{cr}$. This is an effective standard QRM. Despite this, we still provide a possible scheme to observe the superradiant phase of few-body quantum systems in the strong-coupling regime. 
	
In this paper, we consider a negligible cavity dissipation with $\kappa/\omega_{0}<10^{-4}$ and  $\kappa/\omega_{c}^{\prime}<10^{-2}$~\cite{Johansson,RMP,Probing}. When the cavity dissipation is non-negligible, it will produce a shift to the critical coupling strength, which becomes $g_{C}^{\text{diss}}=\frac{1}{2}\sqrt{\frac{\tilde{\omega}_{0}}{\tilde{\omega}_{c}}(\tilde{\omega}_{c}^{2}+\kappa^{2})}$~\cite{dispassive QRM}. For a dissipation rate $\kappa=0.2\omega_{c}^{\prime}=0.002\omega_{0}$ and $\nu/\omega_{0}=0.68$, $\tilde{\omega}_{c}=0.01\omega_{0}$ and $g_{C}^{\text{diss}}\approx1.02g_{C}$. The effective coupling strengths $g_{r}$ and $g_{cr}$ can be larger than $g_{C}^{\text{diss}}$, as shown in Fig.~\ref{Fig2}(c). Hence, it is still effective for our approach to manipulate the steady-state superradiant phase transition of the dissipative QRM.
	
\section{Analysis concerning the two-photon $A^{2}$ term}\label{section7}
It is known that the existence of superradiant phase transition in the cavity-QED system is subject to a no-go theorem, where the squared term of the electromagnetic vector potential (i.e., the $A^{2}$ term) leads to the disappearance of the QPT in the Dicke model~\cite{A2 term}. This no-go theorem also applies to the QPT of the QRM~\cite{Hwang2015Rabi}. In the cavity-QED system, the coefficient $\chi$ in the $A^{2}$ term satisfies $\chi\geq1$ due to Thomas-Reiche-Kuhn sum rule~\cite{no-go theorm}. Thus, the occurrence of the superradiant phase transition is still challenging in the cavity-QED system. However, in the circuit-QED system, $\chi$ could be smaller than 1. Then the superradiant phase transition is recovered with a shifted critical point caused by the $A^{2}$ term~\cite{A2 term,Hwang2015Rabi,no-go theorm,1A2,2A2,3A2}.

In this section, we discuss the validity of omitting the two-photon $A^{2}$ term. The two-photon $A^{2}$ term given by Hamiltonian (\ref{eq:0}) can be written as
\begin{equation}
	H_{A^{2}}=2g_{A^{2}}a^{\dagger}a+g_{A^{2}}(a^{\dagger2}+a^{2}).
\end{equation}
The first term in $H_{A^{2}}$ is absorbed into the Hamiltonian of the standard QRM by replacing the bosonic mode frequency $\omega_{c}$ with the modified frequency $\omega_{c}^{\prime}=\omega_{c}+2g_{A^{2}}$. The second term can be safely neglected with $g\ll\omega_{0},\ g_{A^{2}}\ll2\omega_{c}^{\prime}$. To demonstrate this, in Fig.~\ref{Fig10}, we plot the ratio $g_{A^{2}}/2\omega_{c}^{\prime}$ versus the coupling strength $g$ and the dimensionless coefficient $\chi$. For the dispersion case $\omega_{0}/\omega_{c}=10^{2}$, as shown in Fig.~\ref{Fig10}(a), to satisfy $g_{A^{2}}/2\omega_{c}^{\prime}=0.01$, as indicated by the dashed curve, $g$ and $\chi$ need to be ($g/\omega_{0}\sim0.06,\ \chi\sim0.058$) or ($g/\omega_{0}\sim0.01,\ \chi\sim2$). For the parameters we use in our discussions with $\omega_{0}/\omega_{c}=10^{2}$, $g/\omega_{0}=0.06$, and $\chi\leq0.05$, this ratio is smaller than $1\%$, namely, $g_{A^{2}}/2\omega_{c}^{\prime}<0.01.$ Hence, we can neglect the $A^2$ term in the dispersion limit under the chosen parameters. For the resonance case $\omega_{0}/\omega_{c}=1$, as shown in Fig.~\ref{Fig10}(b), to satisfy $g_{A^{2}}/2\omega_{c}^{\prime}=0.01$, we should have ($g/\omega_{0}\sim0.06,\ \chi\sim5.787$). Therefore, it is also reasonable to omit the $A^{2}$ term for $g/\omega_{0}=0.06$ and $\chi<5$ for the resonance case $\omega_{0}/\omega_{c}=1$. Thus we can omit the $A^{2}$ term by choosing appropriate parameters.

\section{Discussions on the experimental implementation}\label{sectionV111}\label{section8}
In this section, we present some discussions concerning the experimental implementation of our scheme. The key element for the experimental implementation of this scheme is to realize the QRM with the strong coupling between the qubit and bosonic field, as well as the qubit frequency modulation. Here, we present some detailed discussions concerning the realization of the two points. On the one hand, the QPT of the QRM in our scheme occurs when the coupling strength between the qubit and the field is only in the strong-coupling regime, for example $g/\omega_{0}=0.06$. This coupling strength can be realized in various experimental platforms~\cite{3A2,ZhengShiBiao,trapped ions,LiJian,Probing,D. Lv,cqed}. To present some analyses on the current experimental conditions in these systems, the related experimental parameters reported in the circuit-QED system~\cite{ZhengShiBiao} and the trapped-ion system~\cite{trapped ions} are listed in Table \ref{table_two}. The frequency ratio is $\eta\simeq10$ and 25 for the circuit-QED system~\cite{ZhengShiBiao} and the trapped-ion system~\cite{trapped ions}, respectively, which provide the condition for the occurrence of the QPT in QRM.

On the other hand, the qubit frequency modulation is also accessible in several experimental platforms. For a superconducting circuit system, the transition frequency of an Xmon qubit, which couples to a resonator, can be periodically modulated by an ac magnetic flux ~\cite{ZhengShiBiao}. In a trapped-ion system, we can use a linear Pauli trap to confine an ion to simulate the QRM. A pair of counter-propagating Raman laser beams can be used to modulate the transition frequency of the qubit formed by hyperfine states~\cite{trapped ions}. As reported in~\cite{LiJian}, the frequency of the qubit is $\omega_{0}/2\pi=2.62$ GHz and the modulation amplitude is $\xi\nu/2=2\pi\times250$ MHz. One can vary the modulation frequency $\nu/2\pi$ from 0 to 500 MHz, leading to $0\leq\nu/\omega_{0}\leq0.19$ and $\xi>1$. Thus, the modulation parameters reported in experiments are of the same order of magnitude as those used in our scheme. We list the parameters for simulations of our scheme in Table \ref{table_three}. All the parameters are achievable in experiments~\cite{trapped ions,LiJian,ZhengShiBiao,D. Lv}. In addition, we consider the negligible dissipation of the field and qubit. This assumption can also be supported by current experimental technology, based on the fact that the dimensionless decay rates of the field and qubit could be smaller than $10^{-3}$ in the above experimental platforms~\cite{Johansson,RMP,Probing}.

\section{conclusion}\label{sectionV}\label{section9}
We proposed a scheme for generating an anisotropic QRM by applying a qubit frequency modulation to a standard QRM. Here the effective couplings for both the rotating and counter-rotating terms in the obtained anisotropic QRM can be adjusted over a wide range. We calculated the ground-state and excitation energy of this model in terms of the modulation parameters. The phase diagram of the anisotropic QRM in the entire coupling parameter space and the type of the QPT were also obtained. It was shown that the superradiant phase of the QRM can be observed in the strong-coupling regime with the frequency modulation. Furthermore, the two-photon $A^{2}$ term, which could prevent the observation of the QPT in QRM~\cite{Hwang2015Rabi}, can be appropriately neglected in our scheme. We also provided an experimentally accessible scheme to control the QPT in QRM. Our scheme is within current experimental technology~\cite{trapped ions,nuclear,ZhengShiBiao,24.2}. This modulation scheme may also be applicable in other few-body quantum systems. Our work will pave the way for experimentally observing the superradiant phase in other few-body quantum systems.
\begin{acknowledgments}
	J.-F.H. is supported in part by the National Natural Science Foundation of China (Grants No. 12075083 and No.12475016) and the Key Program of Xiangjiang-Laboratory in Hunan province, China (Grant No. XJ2302001). We would like to thank Jie-Qiao Liao for the helpful discussions during the reply to the referees' reports.
\end{acknowledgments}

\appendix
\section{Evaluation of the validity of the rotating-wave approximation}~\label{appC}
To evaluate the validity of the rotating wave approximation used in Hamiltonian (\ref{eq:99}), we examine the time evolution of the fidelity $F(t)=\left|\left\langle \phi\left(t\right)\right|\left.\psi\left(t\right)\right\rangle \right|^{2}$ between the states $\left|\phi\left(t\right)\right\rangle$ and $\left|\psi\left(t\right)\right\rangle$, which are governed by the Hamiltonian (\ref{eq:99}) and Hamiltonian (\ref{eq:4}), respectively. Here, we choose the initial state $\left|\phi\left(0\right)\right\rangle =\left|\psi\left(0\right)\right\rangle =(1/\sqrt{2})\left(\left|g\right\rangle +\left|e\right\rangle \right)\left|\alpha\right\rangle $, where  $\left|\alpha\right\rangle$ is a coherent state of the cavity field.
     
 \begin{table*}
    	\renewcommand{\arraystretch}{1.5}
    	\caption{Experimental parameters reported in the circuit-QED system \cite{ZhengShiBiao} and trapped-ion system \cite{trapped ions}: the transition frequency $\omega_{0}$ of the two-level system, the frequency $\omega_{c}$ of the cavity field, the coupling strength $g$, and the ratio $g/\omega_{0}$.}
    	\label{table_two}
    	\begin{ruledtabular}
    		\begin{tabular*}{\linewidth}{cccccc}
    			Reference & Description & $\omega_{0}/2\pi$ & $\omega_{c}/2\pi$ & $g/2\pi$ & $g/\omega_{0}$\\	
    			\hline
    			\cite{ZhengShiBiao} & Circuit QED & $3.42-10.25$ MHz &$0.34-1.03$ MHz & 0.81 MHz & $0.079-0.237$  \\
    			
    			\cite{trapped ions} & Trapped ion & 50 kHz & 2 kHz& $0-7.1$ kHz & $0-0.142$ \\
    		\end{tabular*}
    	\end{ruledtabular}
 \end{table*}
     
 \begin{table*}[t]
    	\renewcommand{\arraystretch}{1.5}
    	\caption{Analysis of the used parameters in our scheme corresponding to the experimental parameters given in Table \ref{table_two}.}
    	\label{table_three}
    	\begin{ruledtabular}
    		\begin{tabular*}{\linewidth}{cccccc}
    			& &Used parameters  & Circuit QED & Trapped ion\\	
    			\hline
    			Qubit transition frequency	&	$\omega_{0}$  & frequency scale &2$\pi\times$($3.42-10.25$) MHz &2$\pi\times$50 kHz    \\
    			
    			Frequency of the cavity field&$\omega_{c}$&$\omega_{c}/\omega_{0}\simeq0.01$&2$\pi\times$($34.2-102.5$) kHz&2$\pi\times$500 Hz\\
    			
    			Coupling strength	&	$g$&$g/\omega_{0}=0.06$&2$\pi\times$($0.2-0.61$) MHz&2$\pi\times$3 kHz \\
    			
    			Modulation frequency&$\nu$	&	$\nu/\omega_{0}=0.68$&2$\pi\times$($2.32-6.97$) MHz&2$\pi\times$34 kHz\\
    			
    			Dimensionless modulation amplitude	&$\xi$&	$\xi=0-3$&$-$&$-$&\\
    			
    			Modulation amplitude	&	$\xi \nu$&	$\xi \nu/\omega_{0}=0-2.04$&2$\pi\times$($0-20.91$) MHz&2$\pi\times$($0-102$) kHz\\
    		\end{tabular*}
    	\end{ruledtabular}
 \end{table*}
    
\begin{figure}[!b]
	\begin{centering}
		\includegraphics[width=0.49 \textwidth]{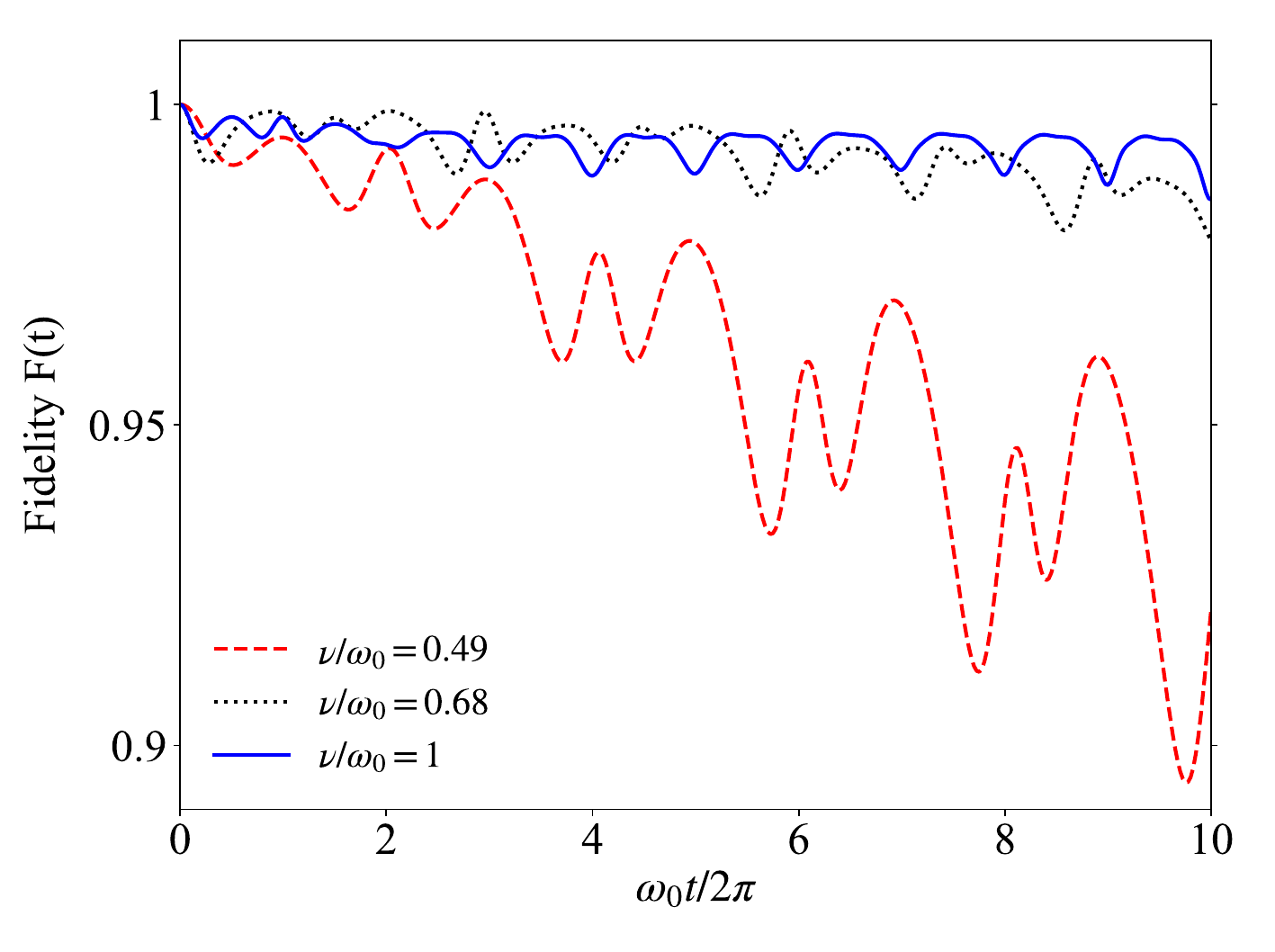}
		\par\end{centering}
	\centering{}\caption{(Color online) Dynamics of the fidelity $F(t)$ for $\nu/\omega_{0}=0.49,\ 0.68, \text{and}\ 1$, respectively. The initial state of the system is $\left|\phi\left(0\right)\right\rangle =\left|\psi\left(0\right)\right\rangle =(1/\sqrt{2})\left(\left|g\right\rangle +\left|e\right\rangle \right)\left|\alpha\right\rangle $ with $\alpha=0.1$. We take the modulation amplitude  $\xi=1.5$. Other parameters are the same as Fig.~\ref{Fig5}.}\label{Fig14}
\end{figure}

We plot the fidelity $F(t)$ versus $t$ for $\nu/\omega_{0}=0.49,\ 0.68, \text{and}\ 1$. As shown in Fig.~\ref{Fig14}, the fidelity can be higher than 0.9 when $\omega_{0}t/2\pi\leq9$ for the used parameters. Specifically, for $\nu/\omega_{0}=0.68,$ which is used in our simulations, the fidelity is higher than 0.97 for $\omega_{0}t/2\pi\leq10$. In particular, the fidelity can be further improved by choosing a suitably larger $\nu$. This result implies that the RWA is valid in our scheme under the chosen parameters.
   
\section{Operators for superradiant phase}~\label{appA}
The form of displacing operator $D_{k}(\alpha)$ ($k$=SX, SP, SXPa, SXPb) for each $k$ phase reads
	
\begin{equation}
	\begin{array}{r@{}l}
		D_{\text{SX}}(\alpha)& {}=e{}^{-\alpha\frac{\partial}{\partial x}},\ \ \ \ \ \ \ \\
		D_{\text{SP}}(\alpha)& {}=e^{-\alpha\frac{\partial}{\partial p}},\ \ \ \ \ \ \ \\
		D_{\text{SXPa}}(\alpha)& {}=e^{-\sqrt{2}\alpha\frac{\partial}{\partial x}},\\
		D_{\text{SXPb}}(\alpha)& {}=e^{-\sqrt{2}\alpha\frac{\partial}{\partial x}}.
	\end{array}
\end{equation}

The new Pauli operators we introduce are
\begin{equation}
	\begin{array}{r@{}l}
		\tau_{x}^{\text{SX}}& {}=\frac{\tilde{\omega}_{0}}{\Omega_{\text{SX}}}\sigma_{x}-\frac{1}{\Omega_{\text{SX}}}\sqrt{2\tilde{\omega}_{0}\tilde{\omega}_{c}}\zeta\left|\alpha_\text{SX}\right|\sigma_{z},\\
		\tau_{y}^{\text{SX}}& {}=\sigma_{y},\ \ \ \ \ \ \ \ \ \ \ \ \ \ \ \ \ \ \ \ \ \ \ \ \ \ \ \ \ \ \ \ \ \ \ \ \ \ \ \ \ \ \ \ \ \ \\
		\tau_{z}^{\text{SX}}& {}=\frac{\tilde{\omega}_{0}}{\Omega_{\text{SX}}}\sigma_{z}+\frac{1}{\Omega_{\text{SX}}}\sqrt{2\tilde{\omega}_{0}\tilde{\omega}_{c}}\zeta\left|\alpha_\text{SX}\right|\sigma_{x}
	\end{array}
\end{equation}
for the SX phase,
\begin{equation}
	\begin{array}{r@{}l}
		\tau_{x}^{\text{SP}}& {}=\frac{\tilde{\omega}_{0}}{\Omega_{\text{SP}}}\sigma_{y}+\frac{1}{\Omega_{\text{SP}}}\sqrt{2\tilde{\omega}_{0}\tilde{\omega}_{c}}\zeta^{\prime}\left|\alpha_\text{SP}\right|\sigma_{z},\\
		\tau_{y}^{\text{SP}}& {}=-\sigma_{x},\ \ \ \ \ \ \ \ \ \ \ \ \ \ \ \ \ \ \ \ \ \ \ \ \ \ \ \ \ \ \ \ \ \ \ \ \ \ \ \ \ \ \ \\
		\tau_{z}^{\text{SP}}& {}=\frac{\tilde{\omega}_{0}}{\Omega_{\text{SP}}}\sigma_{z}-\frac{1}{\Omega_{\text{SP}}}\sqrt{2\tilde{\omega}_{0}\tilde{\omega}_{c}}\zeta^{\prime}\left|\alpha_\text{SP}\right|\sigma_{y}
	\end{array}
\end{equation}
for the SP phase,
\begin{equation}
	\begin{array}{r@{}l}
		\tau_{x}^{\text{SXPa}}& {}=\frac{\tilde{\omega}_{0}}{\Omega_{\text{SXPa}}}\sigma_{x}-\frac{1}{\Omega_{\text{SXPa}}}2g_{c}\left|\alpha_\text{SXPa}\right|\sigma_{z},\\
		\tau_{y}^{\text{SXPa}}& {}=\sigma_{y},\ \ \ \ \ \ \ \ \ \ \ \ \ \ \ \ \ \ \ \ \ \ \ \ \ \ \ \ \ \ \ \ \ \ \ \ \ \ \ \ \ \ \\
		\tau_{z}^{\text{SXPa}}& {}=\frac{\tilde{\omega}_{0}}{\Omega_{\text{SXPa}}}\sigma_{z}+\frac{1}{\Omega_{\text{SXPa}}}2g_{c}\left|\alpha_\text{SXPa}\right|\sigma_{x}
	\end{array}
\end{equation}
for the SXPa phase, and 
\begin{equation}
	\begin{array}{r@{}l}
		\tau_{x}^{\text{SXPb}}& {}=\frac{\tilde{\omega}_{0}}{\Omega_{\text{SXPb}}}\sigma_{x}-\frac{1}{\Omega_{\text{SXPb}}}2g_{r}\left|\alpha_\text{SXPb}\right|\sigma_{z},\\
		\tau_{y}^{\text{SXPb}}& {}=\sigma_{y},\ \ \ \ \ \ \ \ \ \ \ \ \ \ \ \ \ \ \ \ \ \ \ \ \ \ \ \ \ \ \ \ \ \ \ \ \ \ \ \ \ \ \\
		\tau_{z}^{\text{SXPb}}& {}=\frac{\tilde{\omega}_{0}}{\Omega_{\text{SXPb}}}\sigma_{z}+\frac{1}{\Omega_{\text{SXPb}}}2g_{r}\left|\alpha_\text{SXPb}\right|\sigma_{x}
	\end{array}
\end{equation}
for the SXPb phase, respectively.

The form of unitary operator $U_{\text{k}}$ ($k$=SX, SP, SXPa, SXPb) for $k$ phase reads
\begin{equation}
	\begin{aligned}
		U_{\text{SX}}&=\exp[\frac{-i}{\sqrt{2}{\Omega_{\text{SX}}}}\sqrt{\tilde{\omega}_{0}\tilde{\omega}_{c}}( x\tau^{_{\text{SX}}}_{y}/\zeta+\zeta^{\prime}p\tau_{x}^{_{\text{SX}}})], \ \ \ \ \\
		U_{\text{SP}}&=\exp[\frac{i}{\sqrt{2}{\Omega_{\text{SP}}}}\sqrt{\tilde{\omega}_{0}\tilde{\omega}_{c}}(-\zeta x\tau_{x}^{_{\text{SP}}}+ p\tau_{y}^{_{\text{SP}}}/\zeta^{\prime})],\ \ \ \\
			U_{\text{SXPa}}&=\exp\left[\frac{-ig_{c}}{{\sqrt{2}\Omega_{\text{SXPa}}}}\left(\frac{\tilde{\omega}_{0}}{\Omega_{\text{SXPa}}}\tau_{y}^{_{\text{SXPa}}}x-\tau_{x}^{_{\text{SXPa}}}p\right)\right], \\
		U_{\text{SXPb}}&=\exp\left[\frac{-ig_{r}}{{\sqrt{2}\Omega_{\text{SXPb}}}}\left(\frac{\tilde{\omega}_{0}}{\Omega_{\text{SXPb}}}\tau_{y}^{_{\text{SXPb}}}x+\tau_{x}^{_{\text{SXPb}}}p\right)\right].
	\end{aligned}
\end{equation}	

\section{Squeezing parameters}~\label{appB}
The form of the squeezing parameter $r_{p}$ ($p$=N, SX, SP, SXPa, SXPb) is given by
\begin{equation}
	\begin{array}{r@{}l}
			r_{\text{N}}& {}=\frac{1}{4}\ln[(1-\zeta^{2})/(1-\zeta^{\prime^{2}})], \ \ \ \\
			r_{\text{SX}}& {}=\frac{1}{4}\ln[(1-\frac{1}{\zeta^{4}})/(1-\frac{\zeta^{\prime2}}{\zeta^{2}})],\\
			r_{\text{SP}}& {}=\frac{1}{4}\ln[(1-\frac{\zeta^{2}}{\zeta^{\prime2}})/(1-\frac{1}{\zeta^{\prime4}})],\\
			r_{\text{SXPa}}& {}=\infty, \ \ \ \ \ \ \ \ \ \ \ \ \ \ \ \ \ \ \ \ \ \ \ \ \ \ \ \ \\
			r_{\text{SXPb}}& {}=\infty. \ \ \ \ \ \ \ \ \ \ \ \ \ \ \ \ \ \ \ \ \ \ \ \ \ \ \ \   
		\end{array}
\end{equation}
\vspace*{1.8cm}


\providecommand{\noopsort}[1]{}\providecommand{\singleletter}[1]{#1}%

\end{document}